# Signature of Superfluid Density in the Single-Particle Excitation Spectrum of $Bi_2Sr_2CaCu_2O_{8+\delta}$

D.L. Feng,[1*] D.H. Lu,[1] K.M. Shen,[1] C.Kim,[1] H. Eisaki,[1] A. Damascelli,[1] R. Yoshizaki,[2] J.-i. Shimoyama,[3] K. Kishio,[3] G. D. Gu,[4] S. Oh,[5] A. Andrus,[5] J. O'Donnell,[5] J. N. Eckstein,[5] and Z.-X. Shen[1*]

1 Department of Physics, Applied Physics, and Stanford Synchrotron Radiation Laboratory, Stanford University, Stanford, CA 94305, USA.
2 Institute of Applied Physics, University of Tsukuba, Tsukuba, Ibaraki 305, Japan.
3 Department of Applied Chemistry, University of Tokyo, Tokyo 113-8656, Japan.
4 School of Physics, University of New South Wales, Post Office Box 1, Kensington, NSW 2033, Australia.
5 Department of Physics, University of Illinois, Urbana, IL 61801, USA.
* To whom correspondence should be addressed: E-mail: dlfeng@stanford.edu; zxshen@stanford.edu

**We report that the doping and temperature dependence of photoemission spectra near the Brillouin zone boundary of $Bi_2Sr_2CaCu_2O_{8+\delta}$ exhibit unexpected sensitivity to the superfluid density. In the superconducting state, the photoemission peak intensity as a function of doping scales with the superfluid density and the condensation energy. As a function of temperature, the peak intensity shows an abrupt behavior near the superconducting phase transition temperature where phase coherence sets in, rather than near the temperature where the gap opens. This anomalous manifestation of collective effects in single-particle spectroscopy raises important questions concerning the mechanism of high-temperature superconductivity.**

The collective nature of superconductivity manifests itself contrastingly in different techniques. Microwave and muon spin relaxation measurements are inherently sensitive to the collective motion of the condensate, whereas single-electron tunneling spectroscopy and photoemission mainly probe single-particle excitations of the condensate. Hence, these two types of spectroscopies can be used to measure two essential but distinct ingredients of superconductivity: the superfluid density, which characterizes the phase coherence of the Cooper pairs, and the superconducting energy gap, which reflects the strength of the pairing. We report a pronounced departure from this conventional picture on the basis of angle-resolved photoemission spectroscopy (ARPES) data from $Bi_2Sr_2CaCu_2O_{8+\delta}$ (Bi2212).

In Bi2212, a well-known peak and dip feature develops near the superconducting phase transition temperature $T_C$ (1-6) in the ARPES spectra near the Brillouin zone boundary ($\pi/a$,0), where $a$ is the lattice constant, which is set to unity for convenience. This feature has so far been discussed primarily in the context of quasi-particle excitations coupled to many-body collective excitations (3, 7, 8). Here, we show that the doping dependence of the peak intensity exhibits a clear resemblance to the behavior exhibited by the superfluid density $n_s$ and the condensation energy, both of which scale approximately with dopant $x$ in the underdoped regime and saturate or even scale with $A-x$ in the overdoped regime (where $A$ is a constant). The temperature dependence of this peak intensity also shows a resemblance to that of the superfluid density. More important, this peak intensity shows an abrupt behavior near $T_C$, where phase coherence sets in, rather than at $T^*$, the temperature where the pseudogap opens in the underdoped regime (9). It is remarkable that the signature of these collective properties appears in a single-particle excitation spectrum at ($\pi$,0) (the antinode region of a $d$-wave state with maximum gap). This anomalous manifestation of the superfluidity as well as $x$ dependence of many physical quantities contrasts strongly with the conventional Bardeen-Cooper-Schrieffer (BCS) type of picture based on the Fermi liquid. In that picture, the quasi-particle spectral weight $Z$ depends on interactions and the energy gap near the normal state Fermi surface whose volume scales with $1-x$ (counting electrons), rather than on the superfluid density. Instead, these observations agree well with theories that are based on the doped Mott insulator.

We measured ARPES spectra on Bi2212 samples with various doping levels (10). Bi2212 samples are labeled by their $T_C$ with the prefix UD for underdoped, OP for optimally doped, or OD for overdoped (e.g., an underdoped $T_C$ = 83 K sample is denoted UD83). Samples used here include traveling-solvent floating zone-grown single crystals and molecular beam epitaxy (MBE)-grown films. The typical transition width is less than 1 K, except for UD73, which has a transition width of 7 K. Samples with different $T_C$ 's are of similar high quality, as assessed by the measured residual resistivity ratio (RRR), the ratio between the extrapolated resistivity at $T = 0$ K and resistivity at $T = 300$ K. The hole doping level $x$ was determined by the empirical relation $T_C = T_{C,max}[1-82.6(x-0.16)^2]$ (11). $T_{C,max} = 91$ K was used for all the samples because the chemical dopants used in this study (Dy or O) are doped out of the $CuO_2$ plane (12), which changes the doping level but results in much weaker scattering effects than impurities doped in the $CuO_2$ plane (e.g., Zn).

For both the UD83 (Fig. 1A) and the OD84 (Fig. 1B) Bi2212 samples, the spectra near the $(\pi,0)$ region are composed of a peak (open triangles) at low binding energy, followed by a dip (crosses) and a broad hump (solid triangles). These features are clearly distinct below $T_C$ and persist slightly above $T_C$, as shown in insets 2 and 3. A normal-state pseudogap is present in the underdoped sample but not in the overdoped sample (at least for the 110 K spectra) (13). The intensity of the peak in the overdoped sample is much higher than that in the underdoped sample.

To quantify the peak intensity, we focus on the relative peak intensity normalized by the intensity of the entire spectrum. This quantity, which we call the superconducting peak ratio (SPR), eliminates two kinds of artifacts: those attributable to the k-dependent photoemission matrix element, and those that arise because spectra for samples with different dopings are not taken under the exact same conditions. We extracted the peak (as illustrated in Fig. 2A using the spectrum of OD84 at T = 10 K) by fitting the broad hump of the spectra with a five-parameter phenomenological formula,

$$y = a_1 \left( \frac{1}{e^{(x-a_2)/a_3}+1} \right)(1 + a_4(x-a_5)^2) \qquad (1)$$

where the $a_i$'s (i = 1, 2, 3, 4, 5) are the fitting parameters. This formula is simply the product of the Fermi function ($a_3$ is not the temperature) and a parabolic function, and the fit is very robust. The remaining peak then can be fitted by a Gaussian or a Voigt function. The SPR is defined as the ratio between the extracted peak intensity and the total spectrum intensity integrated over [-0.5 eV, 0.1 eV], and this integration window covers the energy scale of the dispersion. We have also used other integration windows, which did not change the qualitative behaviors reported here. Because there are certain subjective factors in the fitting procedure, we have also extracted the peak by subtracting a linear background (Fig. 2B) and an integration (Shirley) background (Fig. 2C), which gave similar results. The SPR values obtained in Fig. 2, A to C, are 0.139, 0.127, and 0.142, respectively. Therefore, errors due to possible subjective factors in the fitting procedure can be estimated to be smaller than 0.01 to 0.02. We chose the phenomenological formula (Eq. 1) for our analysis, because it can smoothly fit the entire hump feature, and we have estimated the errors to be ±1.5% to reflect the subjective uncertainty. This fitting procedure also works well in cases where the peak is small and the dip is weak (Fig. 2, D to F). We stress that the qualitative trends are independent of any given choice of fitting procedure or energy integration window.

Low-temperature spectra (Fig. 3A) were collected at $(\pi,0)$ and at $T = 10$ to $20$ K $T_C$ for different doping levels. The peak is not observed in very underdoped samples, but starts to appear as a small bump in spectrum ud55. Upon further increase of doping, the peak intensity increases until slightly above optimal doping and then decreases in the strongly overdoped regime where the dip disappears. This systematic increase of the peak intensity in the underdoped regime is consistent with earlier data (3, 13, 14). The SPR is plotted against the hole doping level $x$ in Fig. 3B; the SPR increases with doping and reaches a maximum slightly above optimal doping as defined by $T_C$, then decreases in the strongly overdoped regime. We stress here that the ratio of the two parts of the spectra--not just the absolute peak intensity--is changing with doping. Our $k$-dependent data from a few doping levels indicate that the behavior shown in Fig. 3B holds for various points in $k$ space near $(\pi,0)$.

Although scanning tunneling microscopy experiments (12) have found that both the superconducting gap size and the peak intensity are smaller near the scattering centers (impurities or defects), we believe the systematics seen in our "spatially averaged" data are mainly derived from doping for several reasons. First, we do not see a clear correlation between the SPR and residual resistivity, which is a measure of impurity levels. For example, the RRRs of OP91, OD79, and UD30 are ordered as RRR(OP91) < RRR(OD79) ≈ RRR(UD30), whereas the SPRs of these samples are ordered as SPR(UD30) ≪ SPR(OP91) < SPR(OD79) (Fig. 3B). Second, the same behavior of the SPR is observed in samples doped by either Dy or oxygen, which are located at different crystal sites. Third, given the narrow transition widths of most of the samples, it is unlikely that the impurity effects are dominant. Finally, the systematic doping behaviors of the superconducting gap and pseudogap in these samples are consistent with many other measurements.

In comparing the doping and temperature dependence of the SPR with several ground-state quantities related to the superfluidity (Fig. 4), we assume a universality of the properties of the cuprates, because not all the quantities are measured in the Bi2212 system. Our data lend further support to the universality of the doping-dependent behavior of many quantities observed in Bi2212, $YBa_2Cu_3O_7$ (YBCO), and $La_{2-x}Sr_xCuO_4$ (LSCO); because of various experimental difficulties, few systematic studies have been performed on Bi2212 by other techniques. As a function of doping, we find a remarkable resemblance of the SPR (Fig. 4A) to the low-temperature superfluid density as measured by muon spin relaxation (µSR) (Fig. 4B) (15, 16), to the condensation energy from the specific heat ($C_p$) measurements (15), and to the specific heat coefficient jump [$\Delta\gamma_c \equiv \gamma(T_C) - \gamma(120K)$, where $\gamma \equiv C_p/T$] (Fig. 4C) (17). In each of these cases, the physical quantity increases with doping, reaching the maximum slightly above the optimal doping, and then decreases or saturates in the strongly overdoped regime. The decrease of the SPR in the overdoped regime has not been explicitly stated in the published literature.

Upon increasing the temperature, the SPR decreases slowly until about 0.7 $T_C$, then decreases rapidly to zero at a temperature slightly above $T_C$ (Fig. 4D). This temperature dependence suggests that the peak is related to phase coherence and not to the energy gap, because in the underdoped samples, the pseudogap opens well above $T_C$ but the sharp peak shows up only slightly above $T_C$. This temperature dependence of the SPR qualitatively resembles that of $n_s$ as measured by microwave and µSR experiments (Fig. 4E). The microwave and µSR results from the YBCO system are very similar to those from the Bi2212 system, partially justifying our assumption of system universality in the above discussion.

The SPR clearly tracks the superconducting properties measured by microwave and µSR experiments, but there are several discrepancies. First, the $n_s$ measured in these experiments goes to zero at $T_C$, instead of persisting slightly above $T_C$. This discrepancy may be related to the difference in the time scales of the measurements, as photoemission is a much faster probe than microwave measurements or µSR. It has been shown that terahertz optical experiments, a much faster probe than the usual microwave measurements, are sensitive to short-range phase coherence above $T_C$ (18). Second, in the low-temperature regime, the microwave and µSR data exhibit linear temperature dependence, whereas the SPR shows signs of saturation. This may be attributed to the fact that the SPR is obtained from spectra near the $(\pi,0)$ region, where the superconducting gap energy is much larger than the thermal energy. Hence, further lowering the temperature at already low temperatures will not affect the SPR. On the other hand, microwave and µSR experiments measure the overall superfluid density, which is always affected by nodal quasi-particle excitations, and thus lowering the temperature will increase the measured superfluid density by reducing the number of thermally excited quasi-particles. This also suggests that although the SPR is closely related to the superfluid density, it is not an absolute measurement of the superfluid density. For example, although no superconducting peak has been resolved for samples with $T_C \leqslant 40K$ (including Bi2201, LSCO, and Bi2212 in the low doping regime), the superfluid density is clearly not zero in these materials. This is probably also due to smaller superfluid densities and the fact that the measuring temperature (10 to 20 K) is a large fraction of $T_C$ in these systems.

The data reported here raise the following intriguing questions. Why does the SPR scale with $x$ in the underdoped regime? Why does the effect manifest itself most strongly near the $(\pi,0)$ region (i.e., what makes this region so special)? More fundamentally, why is the intensity of the single-particle excitation near the antinode of a *d*-wave superconductor related to the superfluidity? These questions cannot be reconciled within the theoretical framework of superconductivity involving BCS pairing (with either *s*-wave or *d*-wave symmetry) and excitations around a large Fermi surface. The $x$ dependence of the SPR and other quantities requires a fundamental departure from a band-like Fermi surface-based approach in the underdoped regime. Moreover, in a BCS type of picture based on the Fermi liquid concept, the superconducting quasi-particle peak intensity $Z$ should depend on the coherence factor that is related to the energy gap instead of the superfluid density. This theoretical picture contrasts strongly with the fact that the sharp peak in the underdoped regime rises abruptly near $T_C$ rather than $T^*$, where the pseudogap opens. A related problem is that the gap is larger in the underdoped regime while the effect of superconductivity on the photoemission spectra is weaker. The Fermi liquid approach has been extended by attributing the disappearance of the sharp peak above $T_C$ to broadening caused by phase fluctuations (19). However, it is inconsistent with the fact that the integrated peak intensity changes continuously with temperature below $T_C$, whereas the peak width does not change other than by simple thermal broadening over the entire 90 K temperature range, even above $T_C$ (Fig. 4D).

The experimental data reported here are in agreement with the theoretical models based on the doped Mott insulator picture. There have been two classes of theoretical models--resonant valence bond (RVB) gauge theory and the stripe model--that predict the existence of the coherent spectral weight in the single-particle excitation spectrum that is proportional to $x$. The stripe theory envisions a microscopic phase separation that breaks the global two-dimensional (2D) system into local 1D systems of charge stripes and intervening "insulator domains" (20, 21). Electronic structures calculated on the basis of this model reproduce the "flat band" that dominates the spectral weight near the ($\pi$,0) region (22-24), which is also consistent with photoemission data from the statically charge-ordered compound (25). The stripe theory attributes the emergence of the sharp peak below $T_C$ to the phase coherence among the stripes via a 1D to 2D crossover (26). In this picture, the spectral weight of the coherent part of the single-particle excitation spectrum is a monotonic function of the superfluid density rather than of the energy gap (26). This can explain the doping and temperature dependencies of the data, as the number of stripes scales with doping $x$ in the underdoped regime. The RVB gauge theory envisions the superconductivity to be a derived property of the doped Mott insulator, where the pairing force stems from the magnetic interaction already strongly present in the insulator, and doping destroys the residual long-range order and allows RVB pairs to move (27-33). In this picture, an excitation is regarded as a composite of two particles. One of these particles is directly related to the phase of the superconducting order parameter. Following this assumption, this theory naturally gives rise to a coherent quasi-particle whose strength scales with doping $x$ (or phase stiffness) and vanishes above $T_C$. It has also been suggested that the coherent part is most pronounced near $(\pi,0)$ because of the decoupling of the two components of the composite particle in that region (33). In addition to these two classes of theories, there is another model based on the assumption that a quantum critical point (QCP) exists near $x = 0.19$ (15, 34, 35). It suggests that the competing orders on both sides of this QCP cause the nonmonotonic doping behavior of many physical quantities of high-temperature superconductors, such as $n_s$, $\Delta\gamma_c$, and $U$ (Fig. 4, B and C).

Several related issues and obvious questions remain to be explored. One of them is the connection to collective excitations near $(\pi,\pi)$ that are important to quasi-particles near $(\pi,0)$ (3, 7). Of particular relevance here is the well-known neutron $(\pi,\pi)$ resonance mode whose intensity versus temperature is similar to that of the SPR (Fig. 4F) (36). How this mode couples to quasi-particles is currently under discussion (3, 6, 8). Because these are Fermi liquid-based phenomenological theories focused on the issue of the spectral lineshape (3, 6-8), they do not address the key paradox in the data, namely the anomalous correlation between the ARPES peak intensity and the superfluid density in the underdoped regime. Another unresolved issue is the overdoped regime, where the SPR and other physical properties either saturate or decrease. It is still an open question whether this is because the system switches to a "normal" Fermi liquid-like behavior in the overdoped regime, or is due to phase separation (37) or the existence of a QCP at $x$ near 0.19.

In summary, the doping dependence of the SPR at $(\pi,0)$ is found to scale like the collective properties related to the superfluid, particularly the $x$ dependence of the SPR in the underdoped regime. This unexpected manifestation of collective effects in the single-particle excitation spectra cannot be reconciled by models based on Fermi liquid, but rather may be more naturally explained by models based on the doped Mott insulator.

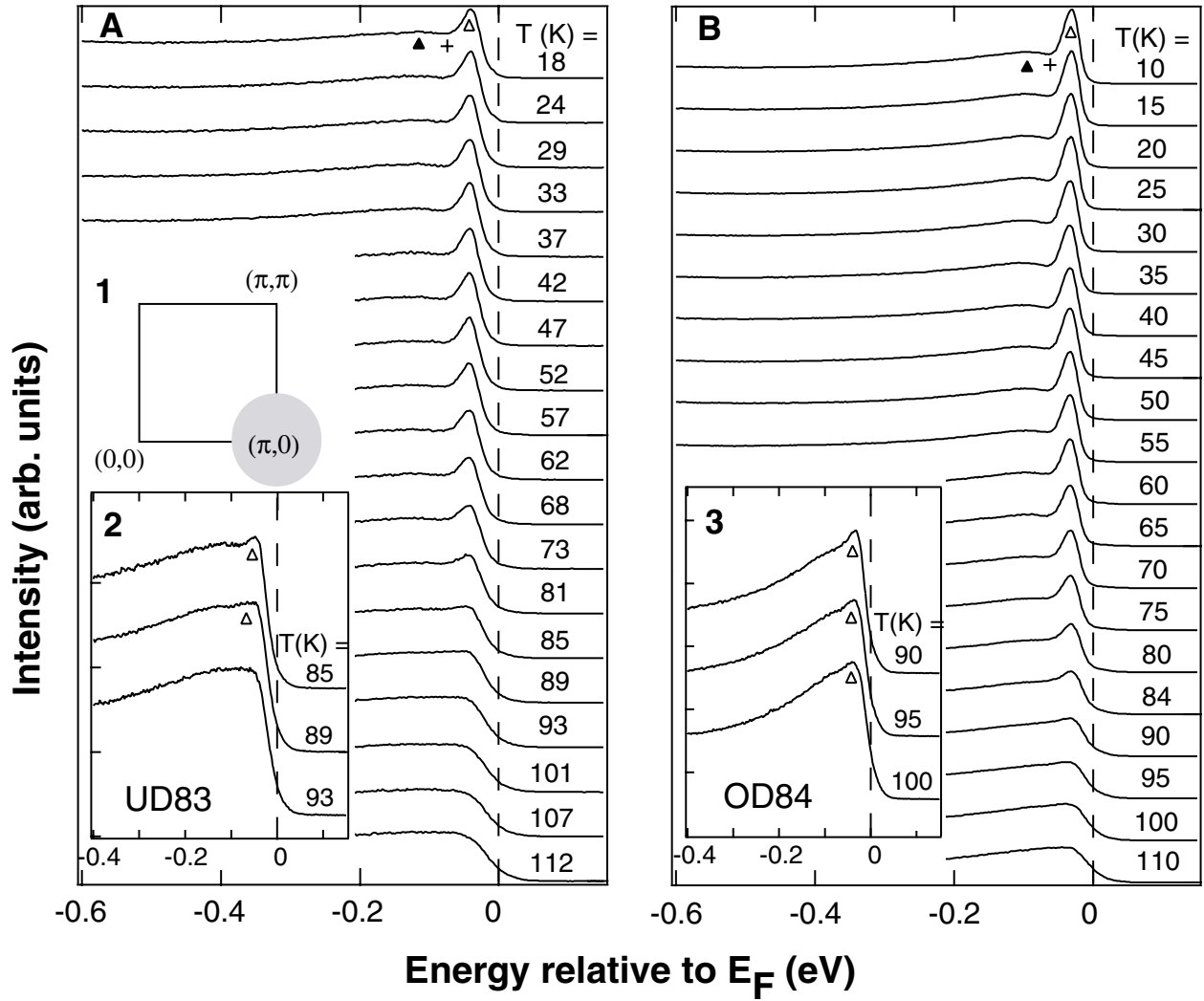

Fig. 1. Temperature dependence of the superconducting state spectra of Bi2212 for (**A**) an underdoped $T_C$ = 83 K sample and (**B**) an overdoped $T_C$ = 84 K sample ($E_F$, Fermi energy). The data are collected near ($\pi$,0) over the shaded circular momentum region in inset **1** to measure the overall relative change of the superconducting peak near the ($\pi$,0) region and to achieve the best signal-to-noise ratio for detailed comparison. Insets **2** and **3** are enlargements of spectra taken above $T_C$; the open triangle markers show that the superconducting peak exists at temperatures slightly above $T_C$.

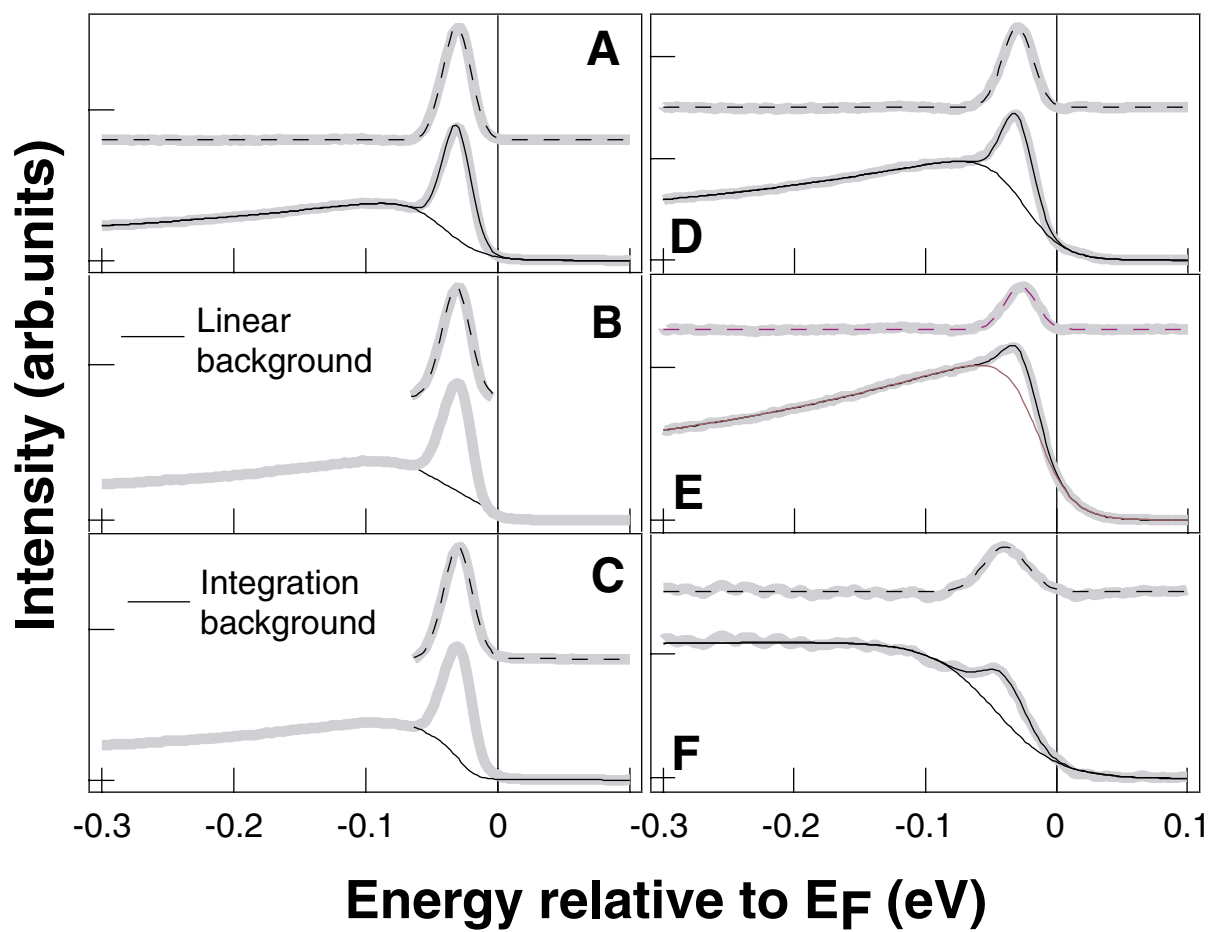

Fig. 2. Various ways of extracting the superconducting peak: (**A**) fitting the hump feature with Eq. 1, (**B**) subtracting a linear background, and (**C**) subtracting an integration background for the spectrum of OD84 at T = 10 K. The same procedure shown in (A) is also applied for (**D**) the spectrum of OD84 at T = 80 K, (**E**) the spectrum of OD84 at T = 90 K, and (**F**) the spectrum of UD55 at T = 15 K [taken from (6)]. In each panel, the lower shaded curves are the raw spectra; the upper shaded curves are the extracted superconducting peaks, which are fitted by a Gaussian function (dashed lines). In (A), (D), (E), and (F), both the fits for the hump feature and the full fits are shown as solid curves.

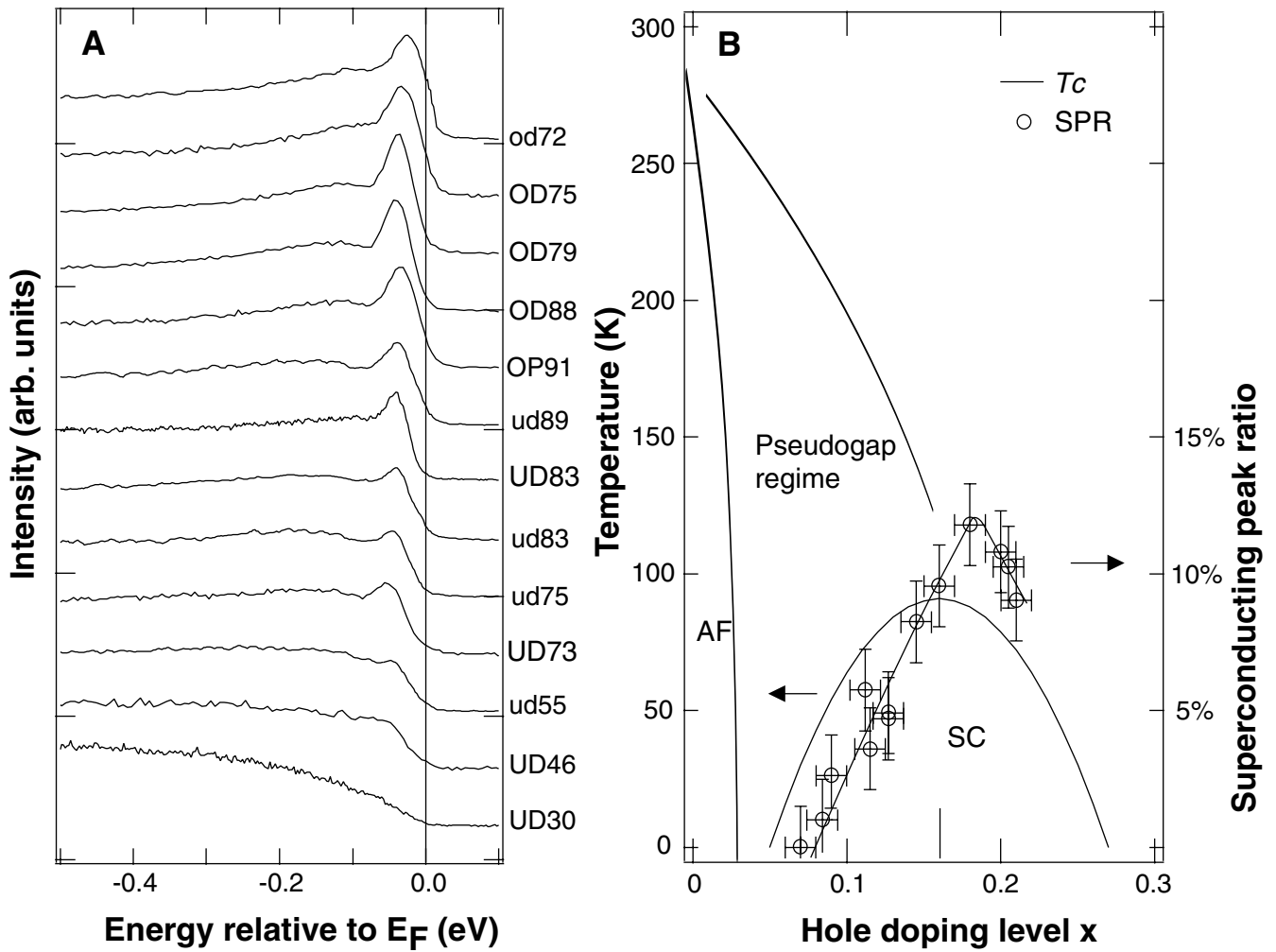

Fig. 3. (**A**) Doping dependence of the superconducting state spectra of Bi2212 at $(\pi,0)$ taken at $T \ll T_C$. Data from samples marked in lowercase are taken from (6). (**B**) The doping dependence of SPR is plotted over a typical Bi2212 phase diagram for the spectra in (A). The solid line is a guide to the eye. Horizontal error bars denote uncertainty in determining the doping level; vertical error bars denote uncertainty in determining the SPR. AF, antiferromagnetic regime; SC, superconducting regime.

D.L. Feng et al. "Science" **280**, p. 277, JULY 2000

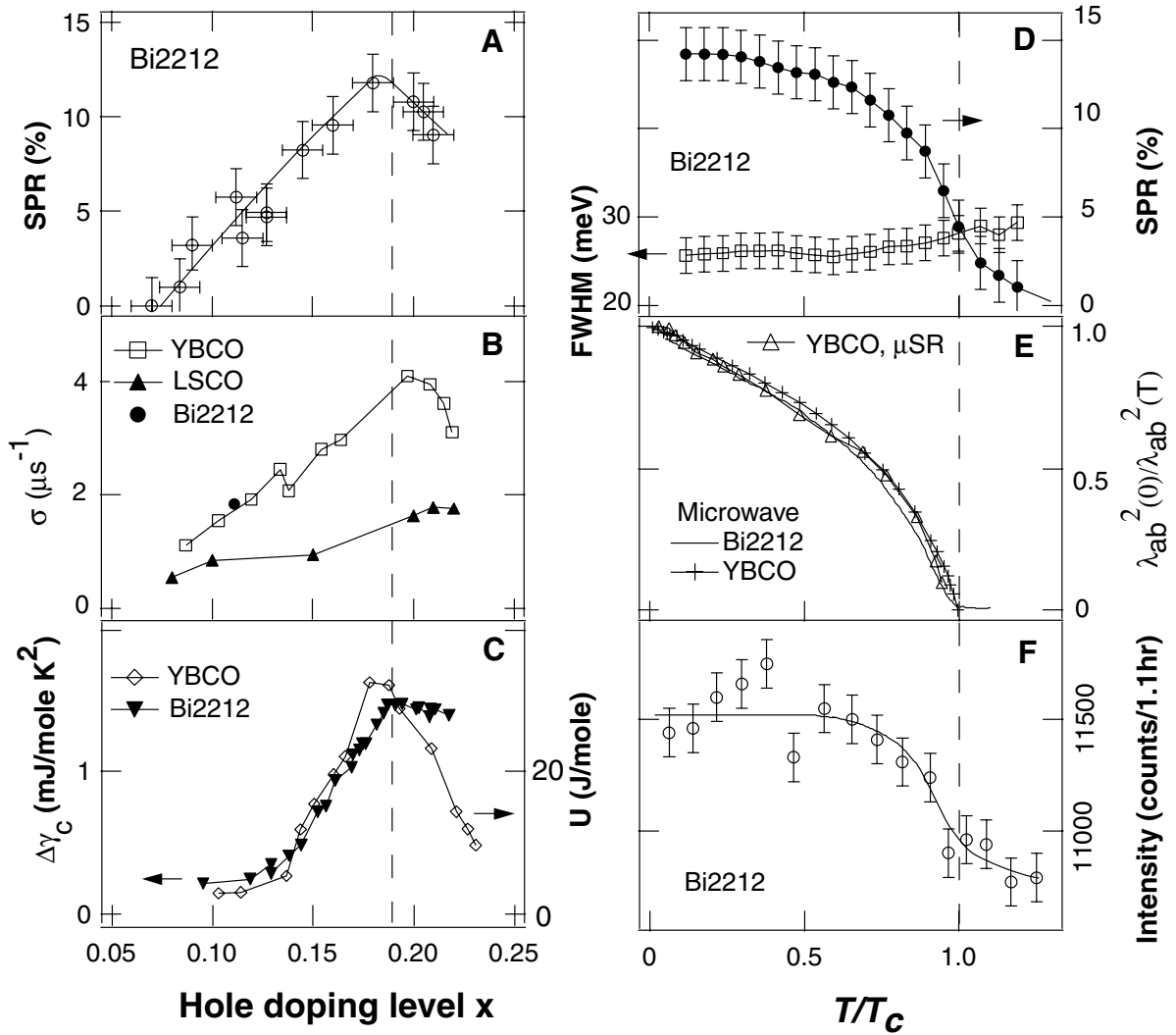

Fig. 4. (**A** to **C**) Low-temperature doping dependence for (A) the SPR of Bi2212 reproduced from Fig. 3B; (B) the SR relaxation rate $\sigma$ (proportional to superfluid density) for YBCO [from (15)], LSCO, and Bi2212 [from (16)]; and (C) the specific heat coefficient jump $\Delta\gamma_c$ of Bi2212 [from (17)] and the condensation energy $U$ of YBCO [from (15)]. The dashed line through (A), (B), and (C) is $x = 0.18$, serving as a guide to the eye. (**D** to **F**) Temperature dependence of (D) the SPR and the peak full width at half-maximum (FWHM) for OD84 data shown in Fig. 1B; (E) $\lambda_{ab}^2(0)/\lambda_{ab}^2(T)$ (proportional to superfluid density), where $\lambda_{ab}$ is the penetration depth, measured by microwave spectroscopy for OP91 Bi2212 [from (38)] and OP93.2 YBCO [from (39)] and by SR for OP93.2 YBCO [from (40)]; and (F) the neutron $(\pi,\pi)$ resonance mode intensity [from (36)] for an OD83 Bi2212.